\documentclass[aps,prl,groupedaddress,superscriptaddress,twocolumn,10pt,longbibliography]{revtex4-2}
\usepackage[utf8]{inputenc}
\usepackage{hyperref}
\usepackage{graphicx}
\graphicspath{{../figures/}}
\usepackage{setspace}
\newcommand{\corauthor}[2]{
    \author{#1}
    \email{#2}
}

\usepackage{siunitx}
\usepackage{braket}
\usepackage{amssymb}
\usepackage{amsmath}
\usepackage{sidecap}
\DeclareSIUnit{\bohrRadius}{\text{$\text{a}_{\textup{0}}$}}

\newcommand{\Ry}{$25\text{D}_{5/2}$ }
\newcommand{\intermediate}{$6\text{P}_{3/2}$ }
\newcommand{\ground}{$5\text{S}_{1/2}$ }
\begin{document}

\title{Engineering long-range molecular potentials by external drive}
\author{Tanita Klas}
\affiliation{Department of Physics and Research Center OPTIMAS, Rheinland-Pfälzische Technische Universität Kaiserslautern-Landau, 67663 Kaiserslautern, Germany}

\author{Jana Bender}
\affiliation{Department of Physics and Research Center OPTIMAS, Rheinland-Pfälzische Technische Universität Kaiserslautern-Landau, 67663 Kaiserslautern, Germany}

\author{Patrick Mischke}
\affiliation{Department of Physics and Research Center OPTIMAS, Rheinland-Pfälzische Technische Universität Kaiserslautern-Landau, 67663 Kaiserslautern, Germany}
\affiliation{Max Planck Graduate Center with the Johannes Gutenberg-Universität Mainz (MPGC), Staudinger Weg 9, 55128 Mainz, Germany}

\author{Thomas Niederprüm}
\affiliation{Department of Physics and Research Center OPTIMAS, Rheinland-Pfälzische Technische Universität Kaiserslautern-Landau, 67663 Kaiserslautern, Germany}

\corauthor{Herwig Ott}{ott@physik.uni-kl.de}
\affiliation{Department of Physics and Research Center OPTIMAS, Rheinland-Pfälzische Technische Universität Kaiserslautern-Landau, 67663 Kaiserslautern, Germany}

\date{\today}

\begin{abstract}
We report the engineering of molecular potentials at large interatomic distances. The molecular states are generated by off-resonant optical coupling to a highly excited, long-range Rydberg molecular potential. The coupling produces a potential well in the low-lying molecular potential, which supports a bound state. The depth of the potential well, and thus the binding energy of the molecule, can be tuned by the coupling parameters. We characterize these molecules and find good agreement with a theoretical model based on the coupling of the two involved adiabatic potential energy curves. Our results open numerous possibilities to create long-range molecules between ultracold ground state atoms and to use them for ultracold chemistry and applications such as Feshbach resonances, Efimov physics or the study of halo molecules.

\end{abstract}

\maketitle



The interaction between individual atoms is determined by the interparticle molecular potentials. They are responsible for the scattering properties and appearance of bound molecular states.
Consequently, a substantial effort is put into properly understanding and controlling interaction potentials, especially in the context of ultracold atomic gases.
A controlled deformation or tailoring of molecular potentials bears great potential for ultracold chemistry applications and few-body physics but is an ambitious task since the molecular potentials are fixed by the atomic properties.
 
Attempts towards potential engineering of low-lying states for scattering processes have attracted considerable attention in the context of magnetic \cite{chin_feshbach_2010} and optical \cite{thomas_experimental_2018} Feshbach resonances.
These have sparked a wide range of advances in many-body physics as the association of ultracold molecules \cite{Regal2003UltracoldMoleculesViaFeshbach,donley_atommolecule_2002}, the formation of molecular Bose-Einstein condensates \cite{greiner_emergence_2003, Jochim2003BECMolecules} and ultracold dipolar molecular systems \cite{Bohn2017ColdMolecules}. 
Macrodimer Rydberg dressing has been demonstrated to create a distance-selective interaction in an optical lattice between two ground state atoms at a micrometer distance \cite{Hollerith2022MacrodimerDressing}.
However, engineering low-lying molecular potentials in such a way, that new bound molecular states emerge, has remained elusive so far.
The control on the molecules' vibrational and rotational state turns potential engineering into a promising platform for ultracold chemistry \cite{liu_bimolecular_2022}. 
\begin{figure}[t]
  \includegraphics[width=\columnwidth]{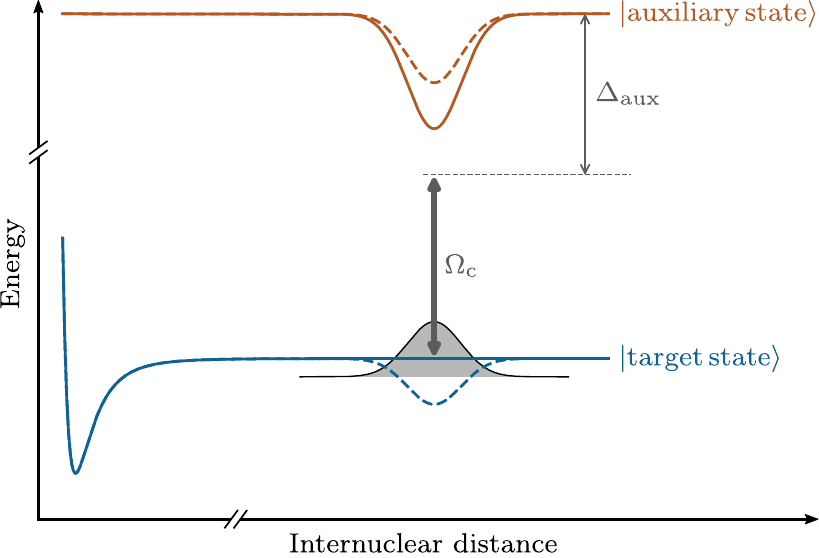}
  \caption{Scheme for molecular potential engineering.
The potential of a molecular target state (solid blue line) is strongly coupled to an excited molecular potential energy curve (solid red line), denoted as auxiliary state.
The off-resonant strong coupling $\Omega_\text{c}$ (grey arrow) with detuning $\Delta_\text{aux}$ modifies both potential curves (blue and red dashed line).
For sufficiently strong coupling, the low-lying potential acquires a potential well that supports a bound state (grey shaded area).
The depth and shape of the potential well can be tuned by the coupling parameters.
For simplicity, the additional atomic lightshift of the involved states is neglected.}
  \label{fig:Fig1}
\end{figure}

Here we demonstrate a generic scheme how to use strong optical coupling to transfer desired characteristics of an auxiliary potential to a target potential.
In particular, we couple an auxiliary potential with a potential well at large interatomic separation to a short-range target potential as shown in Fig.\,\ref{fig:Fig1}.
If, at the interatomic distance of interest, the modulation of the auxiliary potential is large compared to that of the target potential, the character of the auxiliary potential is admixed to the target potential.
The resulting molecular potential inherits the short-range physics from the target state but features an additional potential well at much larger internuclear distances.
If the coupling is strong enough, a bound molecular state emerges.

As recently suggested by Wang and C\^{o}t\'{e} \cite{wang_ultralong-range_2020}, a scheme for potential engineering based on ultralong-range Rydberg molecules is ideally suited to create new potential wells at large interatomic distances in the range of hundreds to thousands of Bohr radii for otherwise only short-ranged potentials.
We apply this generic idea to ultracold Rubidium atoms and manipulate the molecular potential between an atom in the 5S and one in the 6P state (target state) by strong optical coupling to the molecular potential belonging to the 25D Rydberg state (auxiliary state) bound to a ground state atom in the 5S state.
To demonstrate the modification of the potential, we give experimental evidence for the appearance of a molecular bound state in the long-range well of the resulting potential.

By photoassociation spectroscopy, we experimentally find a change in binding energy for different coupling parameters and describe it with a theoretical model.
With their theoretical bond length of $\SI{950}{\bohrRadius}$, the investigated molecules are three orders of magnitude larger than the short-range part of the 5S-6P  potential.
We foresee that such uncommon molecular states and the underlying potential engineering scheme can be employed in the future to functionalize the molecular interaction between ground state atoms in the ultracold regime.

\begin{figure}
    \includegraphics[width=\columnwidth]{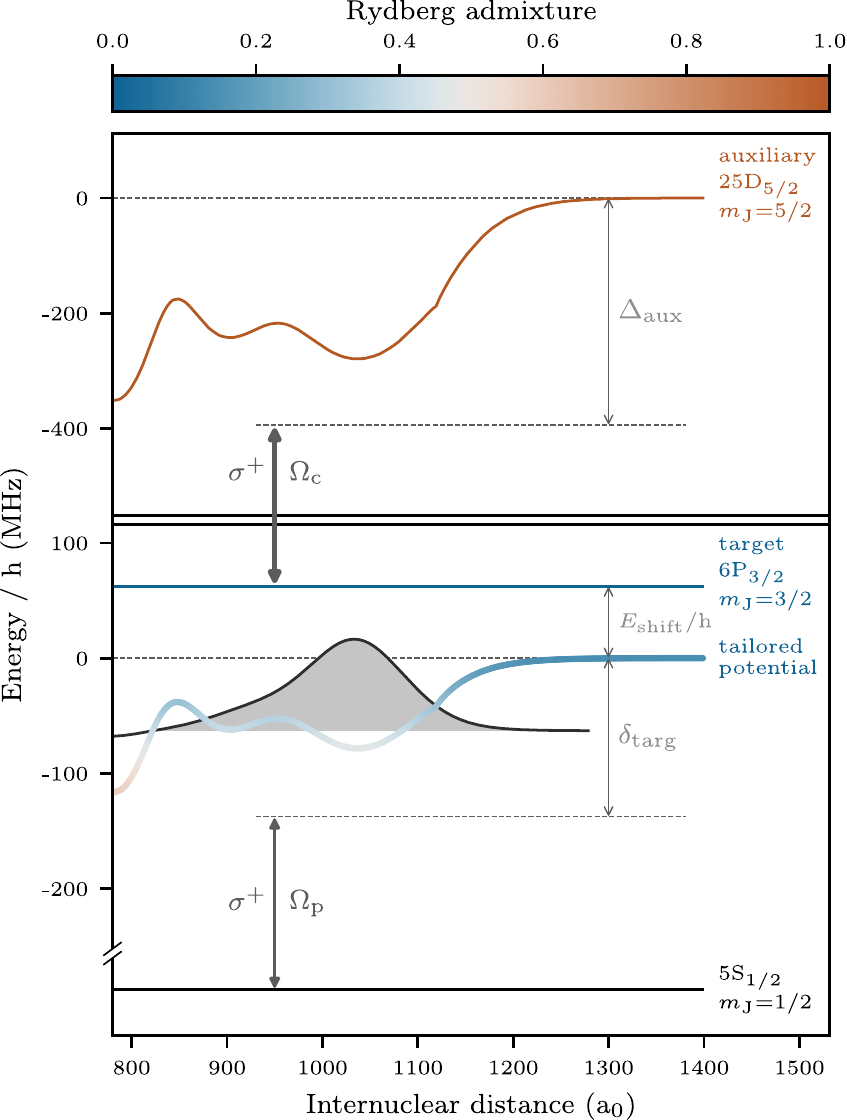}
    \caption{Engineered molecular potential. 
The \intermediate target state with $m_\text{J}=3/2$ (thin blue line) is strongly coupled to the $m_\text{J}=5/2$ sublevel of the auxiliary \Ry Rydberg state (thin red line) with a $\sigma^+$ polarized laser at a detuning $\Delta_\text{aux}=-2\pi\times\SI{300}{\mega\hertz}$ and Rabi frequency $\Omega_\text{c}=2\pi\times\SI{293}{\mega\hertz}$ (grey arrow).
The 5S-6P tailored potential (thick blueish line) is deformed and exhibit a modulated Rydberg state admixture (color code). In particular, the 5S-6P potential acquires a potential well, which supports a vibrational state (grey shaded area), shifted by its binding energy of $-\mathrm{h}\times\SI{62}{\mega\hertz}$.
Photoassociation spectroscopy probes the \intermediate $m_\text{J}=3/2$ state from the \ground $m_\text{J}=1/2$ ground state with a weak, $\sigma^+$ polarized laser and Rabi frequency $\Omega_\text{p}$. Since we treat the Rydberg molecule's binding mechanism separated from the coupling mechanism, we depict the system in a single-atom basis, and consider the 5S atom by the modulation of the Rydberg potential.}
    \label{fig:Fig2}
\end{figure}

To experimentally realize the proposed scheme, the proper choice of the auxiliary Rydberg state is important.
The probability to create such a molecule depends on the distance distribution of the atoms in the underlying ultracold gas and longer bond length and thus higher principle quantum numbers are favored.
A large optical coupling strength as well as the presence of a well isolated and deep potential well in the auxiliary  potential, which allows for larger detunings from the atomic Rydberg state, on the other hand, favor low principle quantum numbers.
Since we drive the coupling transition from the \intermediate state, we can couple to Rydberg D- and S-states from which we generally prefer D-states for their higher ionization rates and stronger Rabi couplings.
Compromising on these considerations, we choose the \Ry state for which the relevant molecular potential is depicted in Fig.\,\ref{fig:Fig2}. 
It provides a double well structure with a depth of about $-\mathrm{h}\times\SI{280}{\mega\hertz}$ at an interatomic distance of $\SI{850}{}$ to $\SI{1100}{\bohrRadius}$ having a sufficient probability to occur at the given density of the atomic gas.
We specifically address its $m_\text{J}=5/2$ Zeeman state to get a defined transition with maximum Rabi coupling from the fully stretched \intermediate $F=3$ $m_\text{F}=3$ state by a $\sigma^+$ transition with a Rabi frequency of $\Omega_\text{c}=2\pi\times\SI{293}{\mega\hertz}$ and varying detunings $\Delta_\mathrm{aux}$ relative to the \Ry Rydberg state.

The confining trap for the underlying sample of about $2\times 10^5$ $^{87}\text{Rb}$ atoms in the \ground $F=2$ $m_\text{F}=2$ ground state is formed by a dedicated beam at \SI{1064}{nm} and the crossed, strong coupling laser at \SI{1030}{\nano\meter}.
\begin{figure*}
  \centering
  \includegraphics[width=\textwidth]{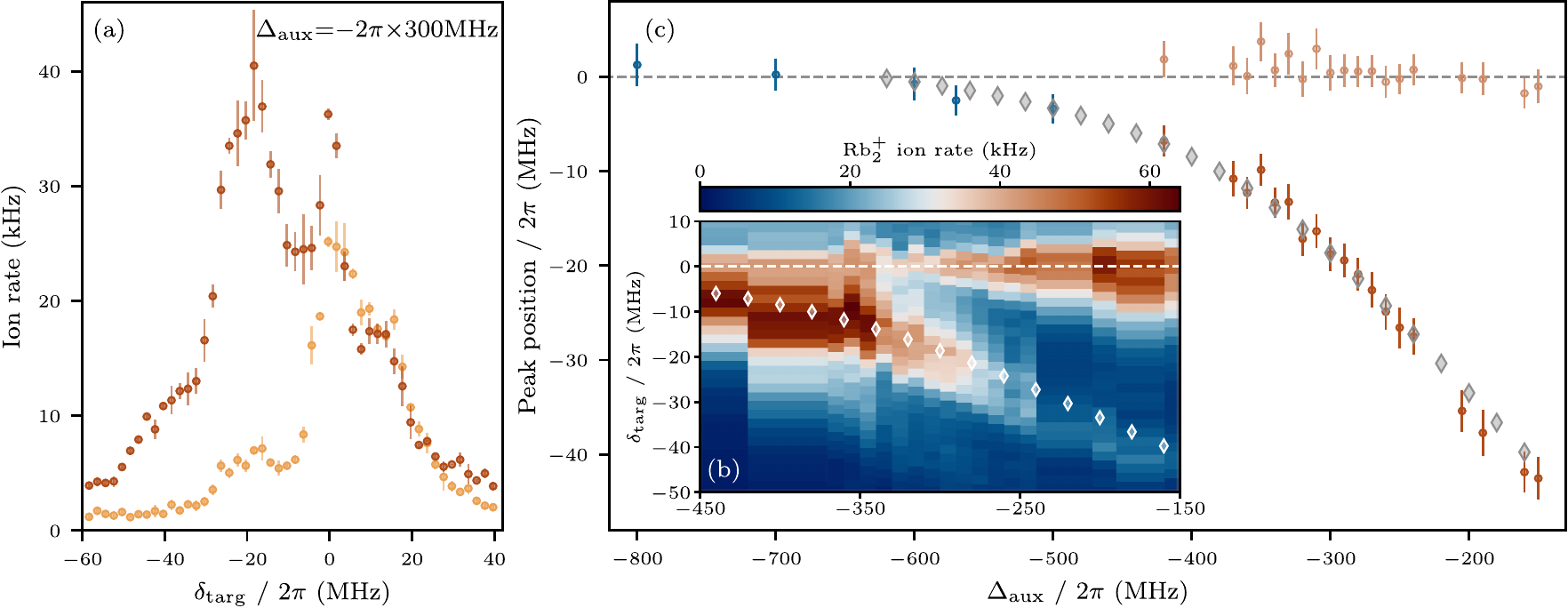}
  \caption{Spectroscopy of the designed 5S-6P molecules for $\Omega_\text{c}=2\pi\times\SI{293}{\mega\hertz}$.
  (a) Atomic Rb$^+$ (orange) and molecular Rb$_2^+$ (red) ions for the detuning of the coupling laser $\Delta_{\rm{aux}}=-2\pi\times\SI{300}{\mega\hertz}$.
Atomic ions stem mainly from the admixture of the Rydberg state to the \intermediate atomic state.
The peak at $\delta_\text{targ} = -2\pi\times\SI{20}{\mega\hertz}$ in the molecular ion signal indicates the presence of a bound 5S-6P state.
$\delta_\text{targ}$ is corrected by the theoretical atomic off-resonant lightshift $E_\mathrm{shift}=-\hbar\Delta_\text{aux}/2-\hbar\sqrt{\Omega_\text{c}^2+\Delta_\text{aux}^2}/2$.
The error bars denote the error of the mean.
(b) We show the Rb$_2^+$ ion rate for different detunings $\Delta_{\rm{aux}}$ of the coupling laser. 
The strong ion rate around $\delta_\text{targ} = 2\pi\times\SI{0}{\mega\hertz}$ (dashed line) represent the
atomic state. 
The additionally appearing line in the map illustrates the presence of 5S-6P molecules. 
The binding energies are convincingly reproduced by calculated bound states in theoretically obtained potential wells that are scaled to 0.39 (white diamonds). The detuning $\Delta_{\rm{aux}}$ at the right edge of the individual measurement
is set in experiment. 
(c) Positions of the atomic (thin red) and molecular peaks (dark red) in the $\text{Rb}_2^+$ signal for different $\Delta_{\rm{aux}}$ are extracted from twp peak Gaussian fits.
For $\Delta_{\rm{aux}}<-2\pi\times\SI{420}{\mega\hertz}$ only a single peak can be identified (blue).
The error bars give the error of the fitted peak position and the statistical error of the lightshift correction as indicated by the atomic peak position variation around $2\pi\times\SI{0}{\mega\hertz}$.
The theoretical peak positions (dashed line and diamonds) are transferred from (b).}
  \label{fig:Fig3}
\end{figure*}
Due to its predominant 5S-6P character, we can photoassociate the new molecular state from the 5S-5S initial state by weakly probing on the 5S to 6P transition. We therefore add a counterpropagating, weak probe beam at \SI{420}{\nano\meter} with $\sigma^+$ polarization, Rabi frequency $\Omega_\text{p}$ and the detuning $\delta_\mathrm{targ}$ relative to the atomic, lightshifted \intermediate state. 
We apply 1000 probe pulses with a length of \SI{10}{\micro\second} while the coupling beam remains at constant, high power for \SI{300}{\milli\second}.
For the detection of the tailored molecules in the 5S-6P target potential we exploit the admixture of Rydberg character and its intrinsic ionization processes into atomic $\text{Rb}^+$ and molecular $\text{Rb}_2^+$ ions.
While $\text{Rb}^+$ ions originate from photoionization by the dipole trap or black body radiation, molecular ions result from associative ionization \cite{niederprum_giant_2015}.
In a small applied electric field the ions are guided to a detector and the two types of ions can be discriminated due to their different time of flight.
Rydberg molecules and atomic Rydberg states differ in their dominant decay channels \cite{schlagmuller_ultracold_2016-1}, so that we expect to find the engineered molecule's signature predominantly in one of the two signals.
We therefore evaluate the two types of ions separately.

Following the probing and coupling scheme depicted in Fig.\,\ref{fig:Fig2}, we measure the spectrum shown in Fig.\,\ref{fig:Fig3}(a).  
The $\text{Rb}^+$ signal shows a strong peak around $\delta_\text{targ}=2\pi\times\SI{0}{\mega\hertz}$  and is thus attributed to atoms in the \intermediate state with admixed Rydberg character.
The experimental signature for the appearance of an engineered 5S-6P molecule is reflected in the $\text{Rb}_2^+$ signal as a strong peak at around $-2\pi\times\SI{20}{\mega\hertz}$.
By systematically changing the coupling between the auxiliary and the target state via the detuning of the coupling laser $\Delta_\text{aux}$ and recording the $\text{Rb}_2^+$ ion rates, we obtain the map shown in Fig.\,\ref{fig:Fig3}(b).
For decreasing $\Delta_\mathrm{aux}$ and thus increasing coupling strength, we find a molecular state branching from the atomic resonance. This demonstrates the continuous deformation of the target state's molecular potential and the tuning capabilities of the bound state.
To extract the position of the molecular and the atomic resonance, we fit a two peak Gaussian function to the $\text{Rb}_2^+$ signal. Their positions are shown in Fig.\,\ref{fig:Fig3}(c). The molecular peak is clearly separated from the atomic resonance at detunings of $\Delta_\mathrm{aux}\geq-2\pi\times\SI{420}{\mega\hertz}$ and its binding energy increases to a maximum of $-\mathrm{h}\times\SI{40}{\mega\hertz}$ for $\Delta_\text{aux}=-2\pi\times\SI{150}{\mega\hertz}$.

To theoretically model our findings we consider again the scheme depicted in Fig.\,\ref{fig:Fig2}, where the difference in coupling strength for the individual Zeeman sub-states is neglected and a single Rabi frequency $\Omega_{\mathrm{c}}$ considered, which couples the \intermediate $F=3$ and the \Ry state with a detuning of $\Delta_{\mathrm{aux}}$. Since the hyperfine interaction in the \intermediate state is smaller than the coupling strength $\Omega_{\mathrm{c}}$, we model the \intermediate state in the uncoupled fine structure basis $\ket{g^*}$:=$\ket{6\mathrm{P}_{3/2}, m_\text{J}, m_\text{I}}$ and the Zeeman states of the \Ry fine structure state $\ket{r}$:=$\ket{25\mathrm{D}_{5/2}, m_\text{J}, m_\text{I}}$ with $I=3/2$ for $^{87}\text{Rb}$.
The Hamiltonian in a single-atom basis of the form $\ket{g^*}\oplus\ket{r}$ is expressed as
\begin{equation}\label{eq:Hamiltonian}
  \hat{H} = \hat{H}_0 + A_\text{HFS}\hat{\vec{I}}\cdot\hat{\vec{J}} + \frac{\hbar}{2}\sum_{g^*,r}\Omega_{g^*r}\ket{g^*}\bra{r}
\end{equation}
where $\hat{H}_{0}$ is the fine structure Hamiltonian, $A_\text{HFS}$ the hyperfine constant, and $\Omega_{g^*r}$ gives the Rabi coupling from the different \intermediate states $\ket{g^*}$ to the Rydberg states $\ket{r}$.
The hyperfine interaction in the Rydberg state is negligible.
Due to the selection rules, the Hilbert space spanned by the $\ket{6\text{P}_{3/2}, 3/2, 3/2}$ and $\ket{25\text{D}_{5/2}, 5/2, 3/2}$ state decouples for $\sigma^+$ polarization of the coupling laser and can be treated as an isolated two-level system.
To ensure that we only probe this subspace in the experiment, we drive a $\sigma^+$ transition from the fully stretched \ground $F=2, m_F=2$ ground state, composed of solely the $m_\text{J}=1/2$ component.
With the spatially varying energy $V_{\text{mol}}(R)$ of the Rydberg molecular potential the system is thus described by
\begin{equation}\label{eq:HamiltonianTwoLevelSysRyMolPot}
  \hat{H}(R)=\hbar\left( \begin{array}{cc}
      0                     & \Omega_\text{c}/2                \\
      \Omega_\text{c}/2 & \Delta_\text{aux}+V_\text{mol}(R)/\hbar \\
    \end{array}\right).
\end{equation}

Fig.\,\ref{fig:Fig2} shows the lower eigenvalues along with the target and auxiliary states of the system as a function of $R$ for typical experimental parameters.
The resulting potential is a mixture of both states and comprises the engineered potential we are aiming for.
At large $R$ the energy difference between the diagonalization result and the target state is given by the AC Stark shift $E_\mathrm{shift}=-\hbar\Delta_\text{aux}/2-\hbar\sqrt{\Omega_\text{c}^2+\Delta_\text{aux}^2}/2$.
From the eigenvectors of Eq.\,\eqref{eq:HamiltonianTwoLevelSysRyMolPot} we extract the local Rydberg state admixture and depict it color coded in Fig.\,\ref{fig:Fig2}.
Via a shooting method we determine the vibrational bound states in the resulting 5S-6P potential and identify the lowest bound state that localizes mostly in the outer double well.
By weighting the local Rydberg admixtures with the vibrational wavefunction, we find an overall Rydberg admixture of about 43 percent for the designed molecule in Fig.\,\ref{fig:Fig3}(a).

The extracted binding energies of the bound states are shown in Fig.\,\ref{fig:Fig3}(c) as grey diamonds. To match the experimentally measured spectra, the calculated molecular potential has been scaled by a factor of 0.39 prior to the calculation of the bound states. 
The reason for this mismatch might stem from neglecting the state mixing in the Rydberg state due to Rydberg molecular interaction, which will lead to less $m_\text{J}=5/2$ character in the auxiliary state and consequently to an $R$-dependent, generally lower Rabi frequency. For the highest Rabi frequency of up to $2\pi\times\SI{293}{\mega\hertz}$, the atom-light coupling might even change the Rydberg-ground state interaction such that our approximation of unperturbed Rydberg molecules becomes increasingly inaccurate.
Apart from this scaling factor, the agreement with the measured spectra is good and the trend of the experimental data as well as the detaching point from the atomic resonance are well reproduced. The spectroscopic result along with the simple two-level explanation demonstrate the ability to deform a target potential to engineer molecular long-range potentials supporting bound states. 


While the short lifetime of the 6P state might limit the applications of our specific realization in ultracold gases, it is important to note that the demonstrated mechanism is generic for all atom pairs, which can be coupled to a Rydberg molecule.
This includes in particular two ground state atoms, which are coupled with a single photon transition to a Rydberg molecular potential.
Because the ground state molecules are stable, only the admixture of the Rydberg molecules shortens the lifetime.
As a consequence, much lower Rabi frequencies are required to create a bound state between the two ground state atoms.
Realistic estimates for the Rabi frequency of $2\pi\times\SI{5}{\mega\hertz}$ and detuning of $-2\pi\times\SI{2}{\mega\hertz}$ suggest a binding energy of about $-\mathrm{h}\times\SI{25}{\kilo\hertz}$ at a molecular lifetime of more than \SI{5}{\milli\second}.
Such conditions would open up plenty of opportunities for ultracold molecular physics.
As the binding energy of the engineered molecules is tunable up to the limit of two free atoms, the creation of halo states \cite{zeller_imaging_2016}, which have a large part of their wave function in the classically forbidden region and thus have an enormous spatial extent, is possible.
Furthermore, the inherent tuning capabilities also directly connect to Efimov physics \cite{naidon_efimov_2017}. Even new realizations of optical Feshbach resonances \cite{thomas_optical_2018} are feasible.

\section*{References}
%

\section{Data availability}
The data that support the plots within this paper and other findings of this study are available from the corresponding author upon request.

\section{Acknowledgements}
We would like to thank R. Côté for helpful discussions. We acknowledges financial support by the DFG within project OT 222/8-1 and the collaborative research center TR185 OSCAR, project B2 (number 277625399). This work was also supported by the Max Planck Graduate Center with the Johannes Gutenberg-Universität Mainz (MPGC).

\section{Author Contribution}
T.K., P.M., and J.B. performed the experiments. T.K. and P.M. analysed the data. T.K. performed the molecular potential calculations and prepared the initial version of the manuscript. H.O. conceived the project. H.O. and T.N. supervised the experiment. All authors contributed to the data interpretation and manuscript preparation.

\section{Competing financial interests}
The authors declare no competing financial interests.


\begin{thebibliography}{15}%
\makeatletter
\providecommand \@ifxundefined [1]{%
 \@ifx{#1\undefined}
}%
\providecommand \@ifnum [1]{%
 \ifnum #1\expandafter \@firstoftwo
 \else \expandafter \@secondoftwo
 \fi
}%
\providecommand \@ifx [1]{%
 \ifx #1\expandafter \@firstoftwo
 \else \expandafter \@secondoftwo
 \fi
}%
\providecommand \natexlab [1]{#1}%
\providecommand \enquote  [1]{``#1''}%
\providecommand \bibnamefont  [1]{#1}%
\providecommand \bibfnamefont [1]{#1}%
\providecommand \citenamefont [1]{#1}%
\providecommand \href@noop [0]{\@secondoftwo}%
\providecommand \href [0]{\begingroup \@sanitize@url \@href}%
\providecommand \@href[1]{\@@startlink{#1}\@@href}%
\providecommand \@@href[1]{\endgroup#1\@@endlink}%
\providecommand \@sanitize@url [0]{\catcode `\\12\catcode `\$12\catcode
  `\&12\catcode `\#12\catcode `\^12\catcode `\_12\catcode `\%12\relax}%
\providecommand \@@startlink[1]{}%
\providecommand \@@endlink[0]{}%
\providecommand \url  [0]{\begingroup\@sanitize@url \@url }%
\providecommand \@url [1]{\endgroup\@href {#1}{\urlprefix }}%
\providecommand \urlprefix  [0]{URL }%
\providecommand \Eprint [0]{\href }%
\providecommand \doibase [0]{https://doi.org/}%
\providecommand \selectlanguage [0]{\@gobble}%
\providecommand \bibinfo  [0]{\@secondoftwo}%
\providecommand \bibfield  [0]{\@secondoftwo}%
\providecommand \translation [1]{[#1]}%
\providecommand \BibitemOpen [0]{}%
\providecommand \bibitemStop [0]{}%
\providecommand \bibitemNoStop [0]{.\EOS\space}%
\providecommand \EOS [0]{\spacefactor3000\relax}%
\providecommand \BibitemShut  [1]{\csname bibitem#1\endcsname}%
\let\auto@bib@innerbib\@empty
\bibitem [{\citenamefont {Chin}\ \emph {et~al.}(2010)\citenamefont {Chin},
  \citenamefont {Grimm}, \citenamefont {Julienne},\ and\ \citenamefont
  {Tiesinga}}]{chin_feshbach_2010}%
  \BibitemOpen
  \bibfield  {author} {\bibinfo {author} {\bibfnamefont {C.}~\bibnamefont
  {Chin}}, \bibinfo {author} {\bibfnamefont {R.}~\bibnamefont {Grimm}},
  \bibinfo {author} {\bibfnamefont {P.}~\bibnamefont {Julienne}},\ and\
  \bibinfo {author} {\bibfnamefont {E.}~\bibnamefont {Tiesinga}},\ }\bibfield
  {title} {\bibinfo {title} {Feshbach resonances in ultracold gases},\ }\href
  {https://doi.org/10.1103/RevModPhys.82.1225} {\bibfield  {journal} {\bibinfo
  {journal} {Rev. Mod. Phys.}\ }\textbf {\bibinfo {volume} {82}},\ \bibinfo
  {pages} {1225} (\bibinfo {year} {2010})},\ \bibinfo {note} {publisher:
  American Physical Society}\BibitemShut {NoStop}%
\bibitem [{\citenamefont {Thomas}\ \emph {et~al.}(2018)\citenamefont {Thomas},
  \citenamefont {Lippe}, \citenamefont {Eichert},\ and\ \citenamefont
  {Ott}}]{thomas_experimental_2018}%
  \BibitemOpen
  \bibfield  {author} {\bibinfo {author} {\bibfnamefont {O.}~\bibnamefont
  {Thomas}}, \bibinfo {author} {\bibfnamefont {C.}~\bibnamefont {Lippe}},
  \bibinfo {author} {\bibfnamefont {T.}~\bibnamefont {Eichert}},\ and\ \bibinfo
  {author} {\bibfnamefont {H.}~\bibnamefont {Ott}},\ }\bibfield  {title}
  {\bibinfo {title} {Experimental realization of a {Rydberg} optical {Feshbach}
  resonance in a quantum many-body system},\ }\href
  {https://doi.org/10.1038/s41467-018-04684-w} {\bibfield  {journal} {\bibinfo
  {journal} {Nature Communications}\ }\textbf {\bibinfo {volume} {9}},\
  \bibinfo {pages} {2238} (\bibinfo {year} {2018})}\BibitemShut {NoStop}%
\bibitem [{\citenamefont {Regal}\ \emph {et~al.}(2003)\citenamefont {Regal},
  \citenamefont {Ticknor}, \citenamefont {Bohn},\ and\ \citenamefont
  {Jin}}]{Regal2003UltracoldMoleculesViaFeshbach}%
  \BibitemOpen
  \bibfield  {author} {\bibinfo {author} {\bibfnamefont {C.~A.}\ \bibnamefont
  {Regal}}, \bibinfo {author} {\bibfnamefont {C.}~\bibnamefont {Ticknor}},
  \bibinfo {author} {\bibfnamefont {J.~L.}\ \bibnamefont {Bohn}},\ and\
  \bibinfo {author} {\bibfnamefont {D.~S.}\ \bibnamefont {Jin}},\ }\bibfield
  {title} {\bibinfo {title} {Creation of ultracold molecules from a fermi gas
  of atoms},\ }\href {https://doi.org/10.1038/nature01738} {\bibfield
  {journal} {\bibinfo  {journal} {Nature}\ }\textbf {\bibinfo {volume} {424}},\
  \bibinfo {pages} {47} (\bibinfo {year} {2003})}\BibitemShut {NoStop}%
\bibitem [{\citenamefont {Donley}\ \emph {et~al.}()\citenamefont {Donley},
  \citenamefont {Claussen}, \citenamefont {Thompson},\ and\ \citenamefont
  {Wieman}}]{donley_atommolecule_2002}%
  \BibitemOpen
  \bibfield  {author} {\bibinfo {author} {\bibfnamefont {E.~A.}\ \bibnamefont
  {Donley}}, \bibinfo {author} {\bibfnamefont {N.~R.}\ \bibnamefont
  {Claussen}}, \bibinfo {author} {\bibfnamefont {S.~T.}\ \bibnamefont
  {Thompson}},\ and\ \bibinfo {author} {\bibfnamefont {C.~E.}\ \bibnamefont
  {Wieman}},\ }\bibfield  {title} {\bibinfo {title} {Atom–molecule coherence
  in a bose–einstein condensate},\ }\href
  {https://www.nature.com/articles/417529a} {\bibfield  {journal} {\bibinfo
  {journal} {Nature}\ }\textbf {\bibinfo {volume} {417}},\ \bibinfo {pages}
  {529}},\ \bibinfo {note} {number: 6888 Publisher: Nature Publishing
  Group}\BibitemShut {NoStop}%
\bibitem [{\citenamefont {Greiner}\ \emph {et~al.}(2003)\citenamefont
  {Greiner}, \citenamefont {Regal},\ and\ \citenamefont
  {Jin}}]{greiner_emergence_2003}%
  \BibitemOpen
  \bibfield  {author} {\bibinfo {author} {\bibfnamefont {M.}~\bibnamefont
  {Greiner}}, \bibinfo {author} {\bibfnamefont {C.~A.}\ \bibnamefont {Regal}},\
  and\ \bibinfo {author} {\bibfnamefont {D.~S.}\ \bibnamefont {Jin}},\
  }\bibfield  {title} {\bibinfo {title} {Emergence of a molecular
  {Bose}-{Einstein} condensate from a {Fermi} gas},\ }\href
  {http://dx.doi.org/10.1038/nature02199} {\bibfield  {journal} {\bibinfo
  {journal} {Nature}\ }\textbf {\bibinfo {volume} {426}},\ \bibinfo {pages}
  {537} (\bibinfo {year} {2003})}\BibitemShut {NoStop}%
\bibitem [{\citenamefont {Jochim}\ \emph {et~al.}(2003)\citenamefont {Jochim},
  \citenamefont {Bartenstein}, \citenamefont {Altmeyer}, \citenamefont {Hendl},
  \citenamefont {Riedl}, \citenamefont {Chin}, \citenamefont
  {Hecker~Denschlag},\ and\ \citenamefont {Grimm}}]{Jochim2003BECMolecules}%
  \BibitemOpen
  \bibfield  {author} {\bibinfo {author} {\bibfnamefont {S.}~\bibnamefont
  {Jochim}}, \bibinfo {author} {\bibfnamefont {M.}~\bibnamefont {Bartenstein}},
  \bibinfo {author} {\bibfnamefont {A.}~\bibnamefont {Altmeyer}}, \bibinfo
  {author} {\bibfnamefont {G.}~\bibnamefont {Hendl}}, \bibinfo {author}
  {\bibfnamefont {S.}~\bibnamefont {Riedl}}, \bibinfo {author} {\bibfnamefont
  {C.}~\bibnamefont {Chin}}, \bibinfo {author} {\bibfnamefont {J.}~\bibnamefont
  {Hecker~Denschlag}},\ and\ \bibinfo {author} {\bibfnamefont {R.}~\bibnamefont
  {Grimm}},\ }\bibfield  {title} {\bibinfo {title} {Bose-einstein condensation
  of molecules},\ }\href {https://doi.org/10.1126/science.1093280} {\bibfield
  {journal} {\bibinfo  {journal} {Science}\ }\textbf {\bibinfo {volume}
  {302}},\ \bibinfo {pages} {2101} (\bibinfo {year} {2003})}\BibitemShut
  {NoStop}%
\bibitem [{\citenamefont {Bohn}\ \emph {et~al.}(2017)\citenamefont {Bohn},
  \citenamefont {Rey},\ and\ \citenamefont {Ye}}]{Bohn2017ColdMolecules}%
  \BibitemOpen
  \bibfield  {author} {\bibinfo {author} {\bibfnamefont {J.~L.}\ \bibnamefont
  {Bohn}}, \bibinfo {author} {\bibfnamefont {A.~M.}\ \bibnamefont {Rey}},\ and\
  \bibinfo {author} {\bibfnamefont {J.}~\bibnamefont {Ye}},\ }\bibfield
  {title} {\bibinfo {title} {Cold molecules: Progress in quantum engineering of
  chemistry and quantum matter},\ }\href
  {https://doi.org/10.1126/science.aam6299} {\bibfield  {journal} {\bibinfo
  {journal} {Science}\ }\textbf {\bibinfo {volume} {357}},\ \bibinfo {pages}
  {1002} (\bibinfo {year} {2017})}\BibitemShut {NoStop}%
\bibitem [{\citenamefont {Hollerith}\ \emph {et~al.}(2022)\citenamefont
  {Hollerith}, \citenamefont {Srakaew}, \citenamefont {Wei}, \citenamefont
  {Rubio-Abadal}, \citenamefont {Adler}, \citenamefont {Weckesser},
  \citenamefont {Kruckenhauser}, \citenamefont {Walther}, \citenamefont {van
  Bijnen}, \citenamefont {Rui}, \citenamefont {Gross}, \citenamefont {Bloch},\
  and\ \citenamefont {Zeiher}}]{Hollerith2022MacrodimerDressing}%
  \BibitemOpen
  \bibfield  {author} {\bibinfo {author} {\bibfnamefont {S.}~\bibnamefont
  {Hollerith}}, \bibinfo {author} {\bibfnamefont {K.}~\bibnamefont {Srakaew}},
  \bibinfo {author} {\bibfnamefont {D.}~\bibnamefont {Wei}}, \bibinfo {author}
  {\bibfnamefont {A.}~\bibnamefont {Rubio-Abadal}}, \bibinfo {author}
  {\bibfnamefont {D.}~\bibnamefont {Adler}}, \bibinfo {author} {\bibfnamefont
  {P.}~\bibnamefont {Weckesser}}, \bibinfo {author} {\bibfnamefont
  {A.}~\bibnamefont {Kruckenhauser}}, \bibinfo {author} {\bibfnamefont
  {V.}~\bibnamefont {Walther}}, \bibinfo {author} {\bibfnamefont
  {R.}~\bibnamefont {van Bijnen}}, \bibinfo {author} {\bibfnamefont
  {J.}~\bibnamefont {Rui}}, \bibinfo {author} {\bibfnamefont {C.}~\bibnamefont
  {Gross}}, \bibinfo {author} {\bibfnamefont {I.}~\bibnamefont {Bloch}},\ and\
  \bibinfo {author} {\bibfnamefont {J.}~\bibnamefont {Zeiher}},\ }\bibfield
  {title} {\bibinfo {title} {Realizing distance-selective interactions in a
  rydberg-dressed atom array},\ }\href
  {https://doi.org/10.1103/PhysRevLett.128.113602} {\bibfield  {journal}
  {\bibinfo  {journal} {Phys. Rev. Lett.}\ }\textbf {\bibinfo {volume} {128}},\
  \bibinfo {pages} {113602} (\bibinfo {year} {2022})}\BibitemShut {NoStop}%
\bibitem [{\citenamefont {Liu}\ and\ \citenamefont
  {Ni}(2022)}]{liu_bimolecular_2022}%
  \BibitemOpen
  \bibfield  {author} {\bibinfo {author} {\bibfnamefont {Y.}~\bibnamefont
  {Liu}}\ and\ \bibinfo {author} {\bibfnamefont {K.-K.}\ \bibnamefont {Ni}},\
  }\bibfield  {title} {\bibinfo {title} {Bimolecular chemistry in the ultracold
  regime},\ }\href {https://doi.org/10.1146/annurev-physchem-090419-043244}
  {\bibfield  {journal} {\bibinfo  {journal} {Annual Review of Physical
  Chemistry}\ }\textbf {\bibinfo {volume} {73}},\ \bibinfo {pages} {73}
  (\bibinfo {year} {2022})}\BibitemShut {NoStop}%
\bibitem [{\citenamefont {Wang}\ and\ \citenamefont
  {Côté}(2020)}]{wang_ultralong-range_2020}%
  \BibitemOpen
  \bibfield  {author} {\bibinfo {author} {\bibfnamefont {J.}~\bibnamefont
  {Wang}}\ and\ \bibinfo {author} {\bibfnamefont {R.}~\bibnamefont {Côté}},\
  }\bibfield  {title} {\bibinfo {title} {Ultralong-range molecule engineering
  via rydberg dressing},\ }\href
  {https://doi.org/10.1103/PhysRevResearch.2.023019} {\bibfield  {journal}
  {\bibinfo  {journal} {Physical Review Research}\ }\textbf {\bibinfo {volume}
  {2}},\ \bibinfo {pages} {023019} (\bibinfo {year} {2020})}\BibitemShut
  {NoStop}%
\bibitem [{\citenamefont {Niederprüm}\ \emph {et~al.}(2015)\citenamefont
  {Niederprüm}, \citenamefont {Thomas}, \citenamefont {Manthey}, \citenamefont
  {Weber},\ and\ \citenamefont {Ott}}]{niederprum_giant_2015}%
  \BibitemOpen
  \bibfield  {author} {\bibinfo {author} {\bibfnamefont {T.}~\bibnamefont
  {Niederprüm}}, \bibinfo {author} {\bibfnamefont {O.}~\bibnamefont {Thomas}},
  \bibinfo {author} {\bibfnamefont {T.}~\bibnamefont {Manthey}}, \bibinfo
  {author} {\bibfnamefont {T.~M.}\ \bibnamefont {Weber}},\ and\ \bibinfo
  {author} {\bibfnamefont {H.}~\bibnamefont {Ott}},\ }\bibfield  {title}
  {\bibinfo {title} {Giant {Cross} {Section} for {Molecular} {Ion} {Formation}
  in {Ultracold} {Rydberg} {Gases}},\ }\href
  {https://doi.org/10.1103/PhysRevLett.115.013003} {\bibfield  {journal}
  {\bibinfo  {journal} {Phys. Rev. Lett.}\ }\textbf {\bibinfo {volume} {115}},\
  \bibinfo {pages} {013003} (\bibinfo {year} {2015})},\ \bibinfo {note}
  {publisher: American Physical Society}\BibitemShut {NoStop}%
\bibitem [{\citenamefont {Schlagmüller}\ \emph {et~al.}(2016)\citenamefont
  {Schlagmüller}, \citenamefont {Liebisch}, \citenamefont {Engel},
  \citenamefont {Kleinbach}, \citenamefont {Böttcher}, \citenamefont
  {Hermann}, \citenamefont {Westphal}, \citenamefont {Gaj}, \citenamefont
  {Löw}, \citenamefont {Hofferberth}, \citenamefont {Pfau}, \citenamefont
  {Pérez-Ríos},\ and\ \citenamefont
  {Greene}}]{schlagmuller_ultracold_2016-1}%
  \BibitemOpen
  \bibfield  {author} {\bibinfo {author} {\bibfnamefont {M.}~\bibnamefont
  {Schlagmüller}}, \bibinfo {author} {\bibfnamefont {T.~C.}\ \bibnamefont
  {Liebisch}}, \bibinfo {author} {\bibfnamefont {F.}~\bibnamefont {Engel}},
  \bibinfo {author} {\bibfnamefont {K.~S.}\ \bibnamefont {Kleinbach}}, \bibinfo
  {author} {\bibfnamefont {F.}~\bibnamefont {Böttcher}}, \bibinfo {author}
  {\bibfnamefont {U.}~\bibnamefont {Hermann}}, \bibinfo {author} {\bibfnamefont
  {K.~M.}\ \bibnamefont {Westphal}}, \bibinfo {author} {\bibfnamefont
  {A.}~\bibnamefont {Gaj}}, \bibinfo {author} {\bibfnamefont {R.}~\bibnamefont
  {Löw}}, \bibinfo {author} {\bibfnamefont {S.}~\bibnamefont {Hofferberth}},
  \bibinfo {author} {\bibfnamefont {T.}~\bibnamefont {Pfau}}, \bibinfo {author}
  {\bibfnamefont {J.}~\bibnamefont {Pérez-Ríos}},\ and\ \bibinfo {author}
  {\bibfnamefont {C.~H.}\ \bibnamefont {Greene}},\ }\bibfield  {title}
  {\bibinfo {title} {Ultracold {Chemical} {Reactions} of a {Single} {Rydberg}
  {Atom} in a {Dense} {Gas}},\ }\href
  {https://doi.org/10.1103/PhysRevX.6.031020} {\bibfield  {journal} {\bibinfo
  {journal} {Phys. Rev. X}\ }\textbf {\bibinfo {volume} {6}},\ \bibinfo {pages}
  {031020} (\bibinfo {year} {2016})},\ \bibinfo {note} {publisher: American
  Physical Society}\BibitemShut {NoStop}%
\bibitem [{\citenamefont {Zeller}\ \emph {et~al.}(2016)\citenamefont {Zeller},
  \citenamefont {Kunitski}, \citenamefont {Voigtsberger}, \citenamefont
  {Kalinin}, \citenamefont {Schottelius}, \citenamefont {Schober},
  \citenamefont {Waitz}, \citenamefont {Sann}, \citenamefont {Hartung},
  \citenamefont {Bauer}, \citenamefont {Pitzer}, \citenamefont {Trinter},
  \citenamefont {Goihl}, \citenamefont {Janke}, \citenamefont {Richter},
  \citenamefont {Kastirke}, \citenamefont {Weller}, \citenamefont {Czasch},
  \citenamefont {Kitzler}, \citenamefont {Braune}, \citenamefont {Grisenti},
  \citenamefont {Schöllkopf}, \citenamefont {Schmidt}, \citenamefont
  {Schöffler}, \citenamefont {Williams}, \citenamefont {Jahnke},\ and\
  \citenamefont {Dörner}}]{zeller_imaging_2016}%
  \BibitemOpen
  \bibfield  {author} {\bibinfo {author} {\bibfnamefont {S.}~\bibnamefont
  {Zeller}}, \bibinfo {author} {\bibfnamefont {M.}~\bibnamefont {Kunitski}},
  \bibinfo {author} {\bibfnamefont {J.}~\bibnamefont {Voigtsberger}}, \bibinfo
  {author} {\bibfnamefont {A.}~\bibnamefont {Kalinin}}, \bibinfo {author}
  {\bibfnamefont {A.}~\bibnamefont {Schottelius}}, \bibinfo {author}
  {\bibfnamefont {C.}~\bibnamefont {Schober}}, \bibinfo {author} {\bibfnamefont
  {M.}~\bibnamefont {Waitz}}, \bibinfo {author} {\bibfnamefont
  {H.}~\bibnamefont {Sann}}, \bibinfo {author} {\bibfnamefont {A.}~\bibnamefont
  {Hartung}}, \bibinfo {author} {\bibfnamefont {T.}~\bibnamefont {Bauer}},
  \bibinfo {author} {\bibfnamefont {M.}~\bibnamefont {Pitzer}}, \bibinfo
  {author} {\bibfnamefont {F.}~\bibnamefont {Trinter}}, \bibinfo {author}
  {\bibfnamefont {C.}~\bibnamefont {Goihl}}, \bibinfo {author} {\bibfnamefont
  {C.}~\bibnamefont {Janke}}, \bibinfo {author} {\bibfnamefont
  {M.}~\bibnamefont {Richter}}, \bibinfo {author} {\bibfnamefont
  {G.}~\bibnamefont {Kastirke}}, \bibinfo {author} {\bibfnamefont
  {M.}~\bibnamefont {Weller}}, \bibinfo {author} {\bibfnamefont
  {A.}~\bibnamefont {Czasch}}, \bibinfo {author} {\bibfnamefont
  {M.}~\bibnamefont {Kitzler}}, \bibinfo {author} {\bibfnamefont
  {M.}~\bibnamefont {Braune}}, \bibinfo {author} {\bibfnamefont {R.~E.}\
  \bibnamefont {Grisenti}}, \bibinfo {author} {\bibfnamefont {W.}~\bibnamefont
  {Schöllkopf}}, \bibinfo {author} {\bibfnamefont {L.~P.~H.}\ \bibnamefont
  {Schmidt}}, \bibinfo {author} {\bibfnamefont {M.~S.}\ \bibnamefont
  {Schöffler}}, \bibinfo {author} {\bibfnamefont {J.~B.}\ \bibnamefont
  {Williams}}, \bibinfo {author} {\bibfnamefont {T.}~\bibnamefont {Jahnke}},\
  and\ \bibinfo {author} {\bibfnamefont {R.}~\bibnamefont {Dörner}},\
  }\bibfield  {title} {\bibinfo {title} {Imaging the {He}$_{\textrm{2}}$
  quantum halo state using a free electron laser},\ }\href
  {https://doi.org/10.1073/pnas.1610688113} {\bibfield  {journal} {\bibinfo
  {journal} {Proceedings of the National Academy of Sciences}\ }\textbf
  {\bibinfo {volume} {113}},\ \bibinfo {pages} {14651} (\bibinfo {year}
  {2016})},\ \bibinfo {note} {\_eprint:
  https://www.pnas.org/doi/pdf/10.1073/pnas.1610688113}\BibitemShut {NoStop}%
\bibitem [{\citenamefont {Naidon}\ and\ \citenamefont
  {Endo}(2017)}]{naidon_efimov_2017}%
  \BibitemOpen
  \bibfield  {author} {\bibinfo {author} {\bibfnamefont {P.}~\bibnamefont
  {Naidon}}\ and\ \bibinfo {author} {\bibfnamefont {S.}~\bibnamefont {Endo}},\
  }\bibfield  {title} {\bibinfo {title} {Efimov physics: a review},\ }\href
  {https://doi.org/10.1088/1361-6633/aa50e8} {\bibfield  {journal} {\bibinfo
  {journal} {Reports on Progress in Physics}\ }\textbf {\bibinfo {volume}
  {80}},\ \bibinfo {pages} {056001} (\bibinfo {year} {2017})},\ \bibinfo {note}
  {publisher: IOP Publishing}\BibitemShut {NoStop}%
\bibitem [{\citenamefont {Thomas}(2018)}]{thomas_optical_2018}%
  \BibitemOpen
  \bibfield  {author} {\bibinfo {author} {\bibfnamefont {O.}~\bibnamefont
  {Thomas}},\ }\emph {\bibinfo {title} {An optical {Feshbach} resonance using
  {Rydberg} molecules}},\ \href
  {https://www.physik.uni-kl.de/fileadmin/ott/diplom-_und_doktorarbeiten/doktorarbeiten/PhD_Oliver_Thomas.pdf}
  {\bibinfo {type} {{PhD} {Thesis}}},\ \bibinfo  {school} {Technische
  Universität Kaiserslautern} (\bibinfo {year} {2018})\BibitemShut {NoStop}%
\end{thebibliography}
\end{document}